\documentstyle{elsart}
\input epsf 
\begin{document}
\newcommand{\lesssim}{\,\vcenter{\hbox{$\buildrel\textstyle<\over\sim$}}}
\newcommand{\gtrsim}{\,\vcenter{\hbox{$\buildrel\textstyle>\over\sim$}}}

\begin{frontmatter}
\title{Casimir Forces at Tricritical Points:\\[4mm]
Theory and Possible Experiments }
\author{U. Ritschel and M. Gerwinski}
\address{Fachbereich Physik, Universit\"at GH Essen, 45117 Essen, Germany}
\maketitle
\begin{abstract}
Using field-theoretical methods and exploiting conformal invariance,
we study Casimir forces at tricritical points
exerted by long-range fluctuations of the order-parameter field.
Special attention is paid to the situation where the symmetry is
broken by the boundary conditions (extraordinary transition).
Besides the parallel-plate configuration, 
we also discuss the geometries of two separate spheres and a single sphere
near a planar wall,
which may serve as a model for colloidal particles immersed in a
fluid. In the concrete case of ternary mixtures
a quantitative comparison with critical Casimir 
and van der Waals forces shows that,
especially with symmetry-breaking boundaries,
the tricritical Casimir force is considerably stronger than the
critical one and dominates also the competing van der Waals force.
Therefore the prospects for an experimental verification
of the Casimir effect at tricritical points should be good.
\end{abstract}
\end{frontmatter}

\section{Introduction}
In 1948 it was discovered by Casimir \cite{Casimir48}
that vacuum fluctuations of the
electromagnetic field give rise to an attractive force
between uncharged conducting capacitor plates. 
Qualitatively speaking, the reason for the Casimir effect 
is that the plates impose boundary conditions on the
zero-point fluctuations of the
electromagnetic field and, as a consequence, the
vacuum energy of the system is changed in a distance-dependent way.
Due to the masslessness of photons, the resulting interaction, the
{\it Casimir force}, is long-ranged. In a recent experiment by
Lamoreaux \cite{lamo} the original result of Casimir was verified to
within $5\%$ accuracy.

In the context of statistical and condensed matter physics it
was pointed out by Fisher and de Gennes \cite{fidege}
that an analogous effect should also exist at or near a {\it critical point}
when the system considered is restricted by boundaries.  
In this case the critical (long-range) fluctuations of
the order-parameter
field play the role of vacuum fluctuations and the critical theory,
in a continuum description, is
formally equivalent to a massless quantum field theory.

In recent years the Casimir effect in statistical
physics was studied theoretically by conformal and perturbative
methods \cite{Bloete86,Affleck,Cardy86,BuXu91,B+E94,Symanzik81,InNi86,Krech91,EiSt,B+E95,eisri} and by means of Monte Carlo simulations\cite{monte}.
However, so far
a direct experimental verification of the statistical Casimir effect
has not been accomplished.
For a review we refer to the book by Krech \cite{Krech94}.
Most of the theoretical efforts
concentrated on the parallel-plate (PP) geometry and on ordinary critical
points. Consider for example a fluid that
is confined between two planar walls, with the lateral extension $M$
and the area $A\propto M^2$,
in a distance $D$ from each other. Further, we assume that $M\gg D$.
In this system the singular part of the free energy at the bulk critical
point $T_c$ asymptotically takes the form \cite{Krech91,Krech94}
\begin{equation}\label{freee}
\frac{F_S}{A\,k_BT_c}\approx M f_{\mbox{\tiny bulk}} +
f_a+f_b + \delta f_{ab}\,,
\end{equation}
where $k_B$ denotes the Boltzmann constant, $f_{\mbox{\tiny bulk}}$ is the
bulk contribution, and $f_a$ and $f_b$
are the surface free energies of surface $a$ and $b$, respectively.
The last term (\ref{freee}) is the Casimir energy. As a consequence of
the scaling invariance of $F_S$ it can be written as
\begin{equation}\label{cas}
\delta f_{ab}=\Delta_{ab}\,D^{-(d-1)}\>.
\end{equation}
It is the contribution to the free energy
that takes into account the interaction between the
walls due to long-range fluctuations. The amplitude
$\Delta_{ab}$ in (\ref{cas}) is a 
universal quantity below the upper critical
dimension $d^*$ and only depends upon the {\em surface 
universality classes} to which the walls belong. For a detailed
analysis of the scaling behavior of $F_S$ away from bulk criticality
we again refer to Refs.\,\cite{Krech91,Krech94}.

How do surfaces affect critical fluctuations?
Now, in the framework of continuum field theory as for example the
$n$-vector model, the surface influence can be
modeled by additional fields
like the surface magnetic field $h_1$
and a local temperature perturbation (surface enhancement), taking into
account the interaction with an adjacent noncritical medium
and a changed coupling strength near the surface
{\it within} the medium, respectively\cite{binder}. 
At $T_c$, where the bulk value of the order parameter---let us call
it $\Phi$---is zero,
the tendency to order near the surface
can be reduced (ordinary transition), increased (extraordinary
transition),
or, as a third possibility,
the surface can be critical as well (special transition).
In the case of both the ordinary and the special transition the
symmetry with respect to $\Phi\to -\Phi$
at the surface is unbroken. In opposition to that, at the extraordinary transition the symmetry is broken explicitly or spontaneously \cite{bray}.
In this case the surface is in an ordered state,
but the bulk, far away from the boundary, is
disordered, the local magnetization $m(z)$ decaying as $\sim z^{-\beta/\nu}$
for {\it macroscopic} distances \cite{binder}.

In the present work, 
we focus our attention on systems at a {\it tricritical
point}, in particular on those with {\it symmetry-breaking} boundaries.
While the existing work on the tricritical Casimir effect
concentrated on symmetry-preserving surfaces \cite{Krech91,Krech94}, there
are several
reasons why also the ``extraordinary'' situation 
at tricritical points should be considered.
It is known that for symmetry-preserving boundary conditions the tricritical
Casimir amplitude is considerably larger than
the corresponding critical amplitude \cite{Krech91,Krech94}.
The same tendency can be expected for symmetry-breaking boundaries.
Additionally, the best candidates
for the experimental verification of the Casimir effect
at tricritical points, ternary 
fluid mixtures, in general lead to a model with symmetry-breaking boundary
conditions.
Consider for instance a ternary mixture near its tricritical point.
The order parameter $\Phi$ of this system is a linear combination
of concentrations of the individual chemical components\cite{lawsa}.
A surface---for example a container wall or the surface of a colloid
particle immersed in the fluid---will generically
favor one of the chemical components, which means that the
$\Phi\rightarrow -\Phi$ symmetry is
broken explicitly by the surface. In turn, in the framework
of lattice or continuum models this is taken into account by a nonzero surface
field $h_1$.
Since $h_1$ is relevant in the sense of the renormalization group
this leads to the scenario of the extraordinary transition
\cite{binder,bray}. The case of $^3$He-$^4$He mixtures, 
more subtle than the one of ternary mixtures, will be addressed in Sec.\,5.

Motivated by an experiment by Ducker et al. \cite{Ducker91},
where the van der Waals force between a spherical (mesoscopic)
particle and a container wall was measured,
we also consider (besides the PP geometry) various
spherical geometries. Recently, it was pointed out
\cite{G+R,B+E95,eisri} that configurations like
two spheres immersed in a critical fluid
and a single sphere near a planar wall can be conformally mapped
to two concentric spheres. Further, due to the rotational symmetry the latter
can be treated on the MF level\cite{G+R}, and, employing
the conformal invariance of $F_S$,
the Casimir force can be calculated in the former
configurations as well \cite{B+E95,eisri}. 

The rest of this article is organized as follows: In Sec.\,2 and 3 the
tricritical Casimir forces are calculated for planar
and spherical geometries, respectively
In Sec.\,4 a quantitative
comparison between van der Waals forces and Casimir forces 
with symmetry-breaking boundaries is carried out. 
In Sec.\,5 possible experiments, especially on ternary
or quaternary mixtures, are discussed, and the issue of
$^3$He-$^4$He mixtures is addressed.

\section{Parallel-Plate Geometry}
 
We first derive the Casimir forces in the
geometry of parallel plates. Since 
$d=3$, the spatial dimension we are interested in, is the upper
critical dimension $d^*$ for tricritical phenomena, 
we have to solve the model on the MF level
and subsequently improve the
result by renormalization-group considerations \cite{WegRie}.
Our starting point is the Landau-Ginzburg
functional for an $n$-component order parameter at the tricritical point
\begin{equation}\label{langin}
{\cal H} = \int_V\,d^3 r\,{\cal L}
\end{equation}
with
\begin{equation}\label{call}
{\cal L}=\frac12 ({\bf \nabla}\vec \Phi)\cdot({\bf \nabla}\vec \Phi) +\frac{g}{6!} \left(\vec \Phi^2\right)^3\>,
\end{equation}
where $V$ denotes that the integration
ranges over the whole critical medium.
The Casimir force in the PP geometry can be
written as \cite{Krech94}
\begin{equation}\label{Casimir}
\frac{1}{A} \frac{d F_s}{d D} = - k_BT_c\, \langle{\cal T}_{zz}\rangle\>,
\end{equation}
where $A$ denotes the area of the plates that, as said above, are assumed
to have a linear extension $M$ much larger than the vertical distance $D$.
In Eq.\,(\ref{Casimir}) and in the following
$T_c$ denotes both critical and tricritical temperatures.
Further, the cartesian stress tensor ${\cal T}_{ij}$
is given by \cite{Brown80}
\begin{equation}\label{stress}
{\cal T}_{ij} =\partial_i\vec \Phi\cdot \partial_j \vec \Phi-
 \delta_{ij} {\cal L} -\frac14 \frac{d-2}{d-1} \left(\partial_i\partial_j
-\delta_{ij}\nabla^2\right) \vec \Phi^2\>,
\end{equation} 
and with (\ref{freee}), (\ref{cas}), and (\ref{Casimir})
one obtains the relation\cite{Krech94}
\begin{equation}\label{relDstress}
\Delta_{ab}=\frac{D^d}{d-1}\,\langle{\cal T}_{zz}\rangle\>
\end{equation}
between the Casimir amplitude introduced in (\ref{cas}) and the $zz$ component
of the stress tensor.

To derive the MF equation from ${\cal H}$,
we let $\vec \Phi$ point in the $1$-direction of the $n$-dimensional
order-parameter space.
Introducing the rescaled field
$\sigma=(g/5!)^{1/4}\Phi_1$, the MF equation reads
\begin{equation}\label{mfequation}
\frac{d^2 \sigma}{d z^2}-\sigma^5=0\>,
\end{equation}
and, after multiplication by $d\sigma/dz$, the first integral
of this equation is found to be
\begin{equation}\label{fint}
I_{\mbox{\tiny PP}}=\left(\frac{d\sigma}{d z}\right)^2-\frac{1}{3}\,\sigma^6\>.
\end{equation}
It only depends on the distance $D$ between the plates and on the
boundary conditions. The latter
can be implemented a posteriori by selecting
solutions that, in the case of symmetry-breaking surfaces,
become singular upon approaching the boundaries $z=\pm D/2$.
In the following
we only consider the situation where $\sigma(z)$ tends 
to $+ \infty$ at both plates and, due to the symmetry,
has a minimum at $z=0$. We
denote this situation as
$\uparrow \uparrow$ boundary conditions,  
as opposed to the situation where for instance
$\sigma$ tends to $\infty$ at one surface and to $-\infty$ at the other,
denoted as the $\uparrow\downarrow$ case.
With the same boundary conditions at both surfaces the resulting
Casimir force is generally {\it attractive}. In the other case, for example
with $\uparrow\downarrow$ boundary conditions, it is {\it repulsive} \cite{Krech94}. This also holds for the spherical geometries
to be discussed in Sec.\,3.

It is straighforward to show
from (\ref{mfequation}) that the first integral is given by
\begin{equation}\label{firsti}
I_{\mbox{\tiny PP}}=-8\sqrt{3}\,\left(\frac{c_6}{D}\right)^3
\end{equation}
with the constant
\begin{equation}
c_6= \int_1^{\infty}\frac{dx}{\sqrt{x^6-1}} = 0.70109\ldots \>.
\end{equation}
Next, there is a close relationship between $I_{\mbox{\tiny PP}}$
and the $zz$-component of the stress tensor (\ref{stress}) that
reads \cite{eisri}
\begin{equation} \label{relti}
{\cal T}_{zz}= \left(\frac{30}{g}\right)^{1/2}  I_{\mbox{\tiny PP}}  \>.
\end{equation}
and can be verified from (\ref{stress}) and (\ref{fint}).
Eventually, the MF (or zero-loop) result for the tricritical Casimir amplitude
is 
\begin{equation}\label{mfcasamp}
\Delta_{\uparrow\uparrow}=-\frac{13.0768...}{g^{1/2}}\>.
\end{equation}
The analogous result for a critical point was derived
in Ref.\,\cite{InNi86}.

The Casimir amplitude in (\ref{mfcasamp}) as it stands still depends
on the arbitrary coupling constant $g$.
In order to remove this dependence, we have to invoke
standard renor\-mali\-zation-group arguments. 
In the scaling limit, the singular part of the
free energy (\ref{freee}) should take the form
\begin{equation}\label{scaling}
{F_S}(M, D, g) = k_BT_c\>{\cal F}(M/D, \bar g(l)),
\end{equation}
where $l\ll 1$ is the spatial rescaling factor and
the running coupling constant $\bar g(l)$ is given by\,\cite{eisdi}
\begin{equation}\label{rung}
\bar g(l) = \frac{240\,\pi^2}{(3n+22)\,
|\mbox{ln}l|}\,\>.
\end{equation}
Other fields (like $\tau$, $h_1$, etc.) that are not
listed in the argument on the left-hand side of
(\ref{scaling}) are assumed
to be adjusted to their fixed-point values and, thus, are unaffected
by the rescaling.

In $d=3-\epsilon$ dimensions
the MF amplitude (\ref{mfcasamp}) would yield the leading 
contribution in a perturbative series, in which 
(after the subtraction of uv divergences)
the renormalized coupling constant could be 
set to its $l\to 0$ value $g^*\propto \epsilon$. 
For $d=d^*$, on the other hand, the coupling must not
be set to its fixed-point value 
$g^*=0$ in (\ref{mfcasamp}). Instead, 
the rescaling parameter $l$ can be replaced
by $a/D$ \cite{brezin},
where $a$ is a microscopic length scale, typically in the range
of a few \AA, as opposed to $D$ which we assume to be 
mesoscopic or macroscopic, of the order of $\mu$m, say.

With (\ref{Casimir}) and (\ref{mfcasamp}) and after
it has been brought into the scaling form with (\ref{scaling}),
the result for the Casimir force reads
\begin{equation}\label{ppcas}
\frac{f^{tc}_{\mbox{\tiny PP}}}{A}\simeq -
0.54\, k_BT_c\, \left(3n+22\right)^{1/2}
\left(\mbox{ln}\,D/a\right)^{1/2} D^{-3}\>.
\end{equation}
Corrections to (\ref{ppcas}) from higher-order terms
in the perturbative series are suppressed logarithmically, i.e.,
by powers of $\mbox{ln}\,D/a$.

At the first glance it appears that the logarithmic term in (\ref{ppcas})
modifies the $D$-dependence of the Casimir force from
$\sim D^{-3}$ to $\sim (\mbox{ln}\,D)^{1/2}|\,D^{-3}$.
That this modification can also be regarded as a correction
and, thus, may be dropped when one only considers the
leading asymptotic behavior, can be seen by rewriting
$\mbox{ln}\,D/a=\mbox{ln}\,D_0/a+\mbox{ln}\,D/D_0$, where $D_0$ is another
macroscopic {\it constant} length scale.
Then the result for the Casimir amplitude is
\begin{equation}\label{casamp}
\Delta_{\uparrow\uparrow}^{tc}=-0.27\,(3n+22)^{1/2}\,\Omega_0\, ,
\end{equation}
where $\Omega_0$ is defined as
\begin{equation}\label{omega}
\Omega_0:=\left(\mbox{ln}\,\frac{D_0}{a}\right)^{1/2}\>.
\end{equation}

As opposed to the situation below $d^*$, where one really finds a
{\it universal} scalar amplitude, at a tricritical point in $d=d^*=3$ we have
additionally a {\it nonuniversal} scalar factor $\Omega_0$,
expressing a dependence of the physics at a macroscopic scale
$D_0$ on the macroscopic scale $a$. 
This phenomenon does not occur in a tricritical system with
both surfaces preserving the symmetry \cite{Krech91}.
Then the leading asymptotic contribution to the Casimir energy
is given by the universal amplitude of the
Gaussian model ($g=0$ in the Eq.\,(\ref{call})). 
In this case only
corrections to scaling depend on the microscopic scale. 
A situation largely analogous to the symmetry-breaking case discussed above
is known from polymers near the $\theta$-point \cite{dupla}.

Even if we do not know $D_0/a$ in (\ref{omega}) in practice,
the dependence on this ratio, as it comes in as the square root
of the logarithm,
turns out to be very weak. For instance when $D_0/a$ is varied
over two orders of magnitude from 100 to 10000 (which should be a reasonable
range from the point of view of the experiments), $\Omega_0$
varies between 2.1 and 3, i.e., it changes
only by about 40$\%$.\\[4mm]
\begin{figure}[t]
\def\epsfsize#1#2{0.63#1}
\hspace*{2cm}\epsfbox{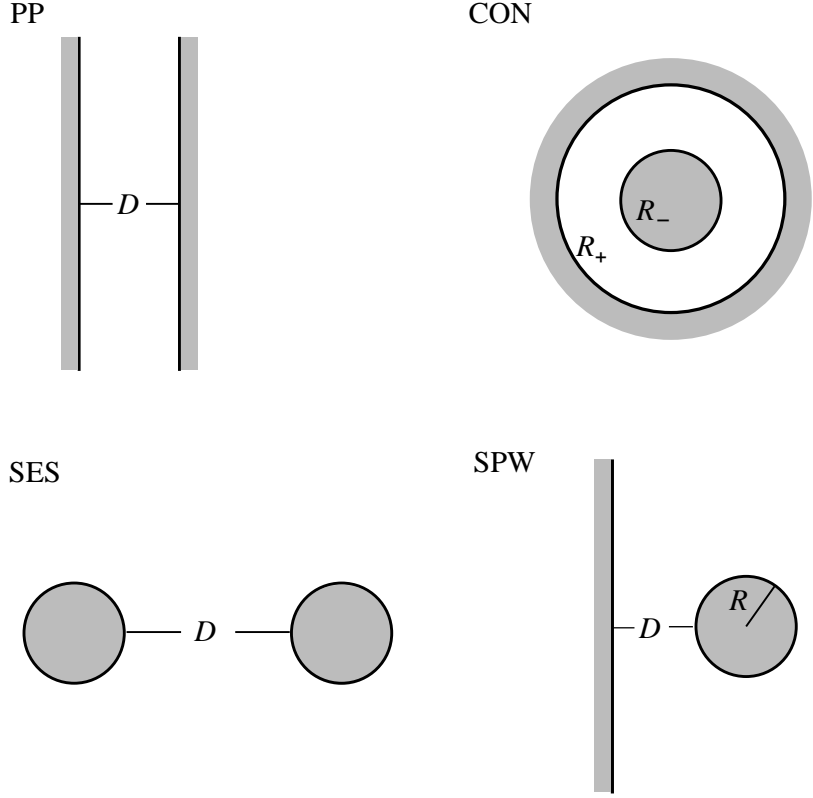}\\[0cm]
\caption{The geometries of parallel plates (PP), concentric spheres (CON),
spheres of equal size (SES), and sphere and planar wall (SPW)
considered in Sec.\,2 and 3. The gray areas represent the noncritical
medium (wall).}
\end{figure}

\section{Spherical Geometries}
Following Ref.\,\cite{eisri} the conformal invariance at a critical
point can be exploited to calculate Casimir forces in
conformally equivalent geometries.
For instance in the highly symmetric geometry of two {\it concentric}
spheres (CON) with radii $R_-$ and $R_+$ of
Fig.\,1, the singular part of the free
energy only depends upon $\rho=R_-/R_+$.  
The derivative of $F_S$ with respect to $\rho$ is given by \cite{eisri}
\begin{equation}\label{deef}
\rho\;\frac{\rm d}{{\rm d}\rho}\,F_S(\rho)=k_BT_c\>S_d\;r_0^d\>\langle {\cal T}_{rr}(r_0)
\rangle_{\mbox{\tiny CON}}\>,
\end{equation}
where ${\cal T}_{rr}$ is the projection of the stress tensor
(\ref{stress}) on the radial direction, $S_d$ denotes the surface area
of the $d$-dimensional unit sphere, and $r_0$ is an inner point
between the two spheres.

The MF equation for the rescaled order-parameter field $\sigma$ that
now only depends on $r$ takes the form
\begin{equation}\label{mfequation2}
\frac{d^2 \sigma}{dr^2}+\frac{d-1}{r}\frac{d \sigma}{dr} -\sigma^5=0.
\end{equation}
This equation is a differential equation of the Emden-Fowler type 
\cite{Leach,G+R,eisri},
and a first integral exists (only) in $d=3$:
\begin{equation}\label{fint2}
I_{\mbox{\tiny CON}}=r^3\frac{d^2 \sigma}{dr^2}+r^2\sigma \frac{d \sigma}{dr} -\frac13
r^3\sigma^6\>.
\end{equation}
Like in the PP case we are interested in $\uparrow
\uparrow$ boundary conditions, i.e. where $\sigma$ becomes
infinite at both the inner and the outer sphere, respectively. 
In this case there exists a
relation between $\rho$ and
$I_{\mbox{\tiny CON}}$ that can be expressed in the form
\begin{equation}\label{rhovoni}
\rho=\exp\left[-\sqrt{\frac{3}{pq}}\,F(\alpha,k)\right]\>,
\end{equation}
where $F(\alpha,k)$ is the elliptic integral of the first kind \cite{Gradst},
and the arguments are
\begin{eqnarray}\label{const} 
\alpha&=&2\,\mbox{arctan}\left(\sqrt{\frac{q}{p}}\,\right)\nonumber\\[2mm]
k& =& \sqrt{\frac{(p+q)^2-a^2}{4pq}}
\end{eqnarray}
with
\begin{eqnarray}
p & =& \sqrt{b^2+\frac94 a^2}\nonumber\\[2mm]
q &=& \sqrt{b^2+\frac14 a^2}\nonumber\\[2mm]
a&=& A_++A_-\nonumber\\[2mm]
b&=&\frac{\sqrt{3}}{2}\left(A_+-A_-\right),
\end{eqnarray}
and the dependence of $A_{\pm}$ on $I_{\mbox{\tiny CON}}$ is given by
\begin{equation}\label{deponi}
A_{\pm}=\left(-\frac{3}{2} 
I_{\mbox{\tiny CON}}  \pm\frac18
\sqrt{1+144 I_{\mbox{\tiny CON}}^2}\right)^{1/3}\>.
\end{equation}
Moreover, analogously to (\ref{relti})
the relation between the first integral and the stress tensor reads
\begin{equation}\label{stri}
r_0^3\, T_{rr}(r_0) = \left(\frac{30}{g}\right)^{1/2}
\! I_{\mbox{\tiny CON}}\>
\end{equation}
such that on the MF level the Casimir force in (\ref{deef})    
can be obtained as a function of $\rho$.

Now, the concentric geometry is conformally equivalent to various 
other geometries \cite{eisri}. For example by inversion at an inner
point in between the concentric spheres the system is mapped to a geometry
where two separate spheres are immersed in a surrounding
(critical) medium.
If the point of inversion lies at the geometric mean $(R_-\,R_+)^{1/2}$,
one obtains {\it spheres} of {\it equal size} (SES). 
Likewise, inversion at one of the radii, $R_-$ or $R_+$, 
generates a geometry in which a single {\it sphere}
is placed in a given distance from a {\it planar wall} (SPW).
All the geometries treated in the present work are depicted in Fig.\,1.

At (and above) the upper critical dimension conformal invariance
does not hold anymore \cite{CardyRev}.
However, on the MF level the field theory described by (\ref{call})
is conformally invariant just in $d=3$. 
Further, since we
are only interested in boundary conditions that correspond to
renormalization-group fixed points, conformal invariance is 
not affected by the boundaries either\cite{CardyPaper}.
Hence, the MF approximation to the Casimir force,
obtained for instance in the CON geometry, can
directly be translated to other conformally equivalent geometries 
(like SPW or SES).
 
As discussed in more detail in Ref.\,\cite{eisri}, 
$F_S$ depends upon certain
invariant cross ratios of the geometric dimensions \cite{CardyRev},
a convenient parameter to quantify the dependence on the geometry being
\begin{equation}\label{kappa}
\kappa = \left\{ \begin{array}{ll}
\frac12 \left(\rho+{1/\rho}\right) & \quad\mbox{for CON}   \\[2mm]
 D^2/(2R^2)+2D/R+1  & \quad \mbox{for SES}\\[2mm]
D/R+1 &\quad \mbox{for SPW}
\end{array}\right.\>.
\end{equation}
Thus for instance the force in the SPW geometry
derived from the known result of the CON geometry, is given by  
\begin{equation}\label{transl}
f^{tc}_{\mbox{\tiny SPW}}=-\frac{d F_S}{d D} = 
-\frac{dF_S}{d\rho} \frac{d\rho}{d D}\>,
\end{equation}
where the first factor on the right-hand side can be derived
from (\ref{deef}), and $d\rho/dD$ can be calculated from
(\ref{kappa}). With (\ref{stri}) we obtain the MF result
\begin{equation}\label{MFSPW}
f^{tc}_{\mbox{\tiny SPW}} = \,k_BT_c \,\left(\frac{30}{g}\right)^{1/2}
\frac{I_{\mbox{\tiny CON}}(\rho)}{(D^2+2DR)^{1/2}}\>.
\end{equation}

As the next step, in the new geometry, the MF result has to be
improved by means of scaling arguments, much in 
the same way as discussed in Sec.\,2 for the PP geometry.
The renormalization-group improvement
again yields a nonuniversal factor
$\Omega_0$ (see Eq.\,(\ref{omega})), 
where now distance scales of the new geometry 
(SPW or SES) enter.

Eventually, the results for the Casimir force in the SPW and SES
geometries are given by
\begin{equation}\label{SPW}
f^{tc}_{\mbox{\tiny SPW}} = k_BT_c
(6n+44)^{1/2} \,\Omega_0 \,
\frac{I_{\mbox{\tiny CON}}(\rho)}{(D^2+2DR)^{1/2}}\>
\end{equation}
and
\begin{equation}\label{SES}
f^{tc}_{\mbox{\tiny SES}} = k_BT_c (6n+44)^{1/2}\, \Omega_0 \,
\frac{2\, I_{\mbox{\tiny CON}}(\rho)}{(D^2+4DR)^{1/2}}\>,
\end{equation}
respectively, where $\rho=\rho(D)$ can be obtained from
(\ref{kappa}).
For the concrete example of a ternary mixture 
with $T_c=300\,$K,
spheres with a radius $R=1\,\mu$m, and 
with the nonuniversal factor set to$\Omega_0=2$,
the results are plotted in Figs.\,1 and 2 and compared with
the van der Waals forces.

\section{Comparison with the critical Casimir effect
and van-der-Waals forces}
For a quantitative comparison between
the tricritical and the critical
Casimir
effect and the van der Waals forces let us first
consider the PP geometry and a system with
an Ising-like order parameter ($n=1$).
Then from (\ref{casamp}) we find for the tricritical
Casimir amplitude $\Delta_{\uparrow\uparrow}^{tc}=-1.35\,\Omega_0$,
where the nonuniversal factor $\Omega_0$ was defined in (\ref{omega}).

In a recent preprint \cite{Krech97}, Krech obtained
$\Delta_{\uparrow\uparrow}^{c}\simeq -0.35$
for the Casimir amplitude of a critical Ising-like
system in $d=3$. Hence, taking into account that
$\Omega_0\gtrsim 2$, the tricritical Casimir amplitude is about
seven times larger then the corresponding critical amplitude,
and, at least as far as the magnitude of the force is concerned,
tricritical points should be better suited for
the experimental verification of the Casimir effect than
critical points at comparable temperatures.    

How does the Casimir effect compare with the van der Waals force?
The result for the van der Waals force in the PP geometry
is given by \cite{israel}
\begin{equation}\label{ppvdw}
\frac{f_{\mbox{\tiny PP}}^{vdW}}{A}=-\frac{H}{6\pi}\,\frac{1}{D^3}\>
\end{equation}
where $H$ is the Hamaker constant, a material-dependent
quantity that in most systems
lies somewhere between $10^{-19}$J and $10^{-21}$J \cite{israel}.
For the sake of concreteness, let us set the Hamaker constant to a
typical value of about $H= 10^{-20}$J.
Taking then the ratio between tricritical
Casimir and van der Waals force we find
\begin{equation}\label{ratio}
\frac{f_{\mbox{\tiny PP}}^{tc}}{f_{\mbox{\tiny PP}}^{vdW}}
\simeq 0.07 \,\Omega_0\, T_c/\mbox{K}\>.
\end{equation}
Hence, even if the tricritical temperature
were about $1\,$K---in Helium mixtures it is
$0.867\,$K--- 
and we set
$\Omega_0\simeq 2$ as before, the
Casimir force amounts to
more than $10\%$ of the van der Waals force. This is different
if symmetry-preserving boundaries
are considered; in this case
the Casimir force is merely a $1\%$ effect \cite{Krech91}.
For ternary mixtures, where usually systems are studied with
$T_c$ near room temperature,
the Casimir force becomes even considerably
stronger than the van der Waals attraction.
\begin{figure}[htbp]
\def\epsfsize#1#2{0.8#1}
\hspace*{2cm}\epsfbox{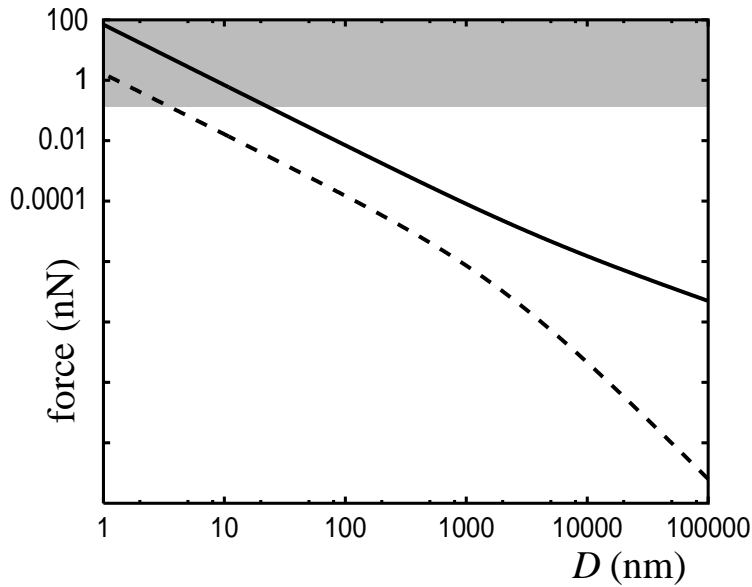}\\[0cm]
\caption{The tricritical Casimir force (solid) and the
van der Waals force (dashed) in SPW geometry of Fig.\,1
for $T_c=300\,$K, particle radius $R=1\,\mu$m and Hamaker constant
$H=10^{-20}\,$J. The grey shaded area indicates the regime 
with force $f\gtrsim 0.2\,$nN where
direct measurements should be feasible (cf. Ref.\,[20].)}
\end{figure}
 
As the next example let us consider the case SPW in Fig.\,1.
In this geometry direct measurements
of the van der Walls force by means of an atomic force microscope were
reported by Ducker et al. \cite{Ducker91}, and it should be feasible
to carry out
a similar measurement with a critical
or tricritical fluid.
For the SPW geometry an approximative expression for the 
nonretarded van der Waals force \cite{israel}  
\begin{equation}\label{vdWspw}
f_{\mbox{\tiny SPW}}^{vdW}= -\frac{2H}{3}\,\frac{R^3}{(D+2R)^2D^2}
\end{equation}
can be obtained.  
A direct comparison between (\ref{vdWspw}) and our result
(\ref{SPW}) is carried out in Fig.\,2. 
Again, $H$ was set to the typical value
of $10^{-20}$ J. For the Casimir force
(solid curve)
we assumed that $T_c=300\,$K and $\Omega_0=2$. For large
distances $D\gg R$, the Casimir force decays more slowly
than the van der Waals result (dashed line), the former
behaving as $\sim D^{-3/2}$ and the latter as $D^{-4}$. For small
(but still macroscopic) distances
distances $a\ll D\ll R$, both the Casimir and the van der Waals force show
the same $D$-dependence $\sim D^{-2}$. The reason for this coincidence
is that this limit
is closely connected to the PP geometry\cite{eisri}, the
ratio of amplitudes being given by (\ref{ratio}).
The grey shaded area in Fig.\,2 shows the regime with
forces $\gtrsim 0.2\,$nN,
where the measurement with an atomic
force microscope, like the one reported
in Ref.\,\cite{Ducker91}, should be possible.
\\[4mm]
\begin{figure}[htbp]
\def\epsfsize#1#2{0.8#1}
\hspace*{2cm}\epsfbox{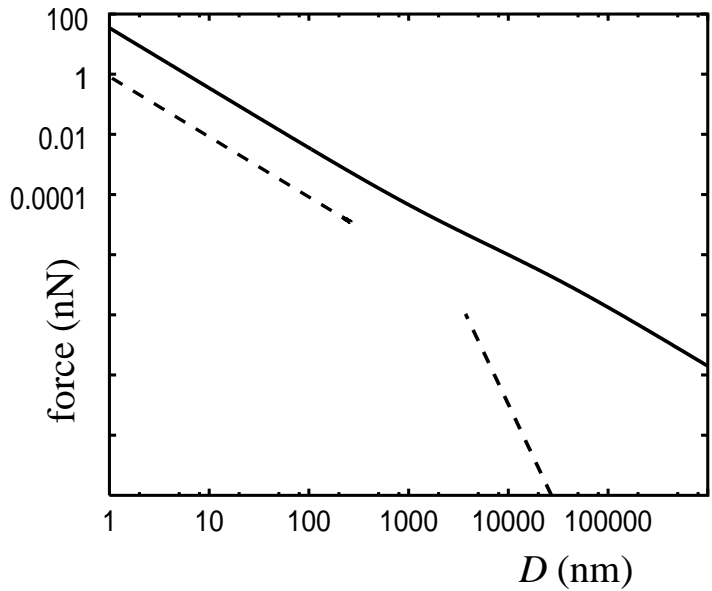}\\[0cm]
\caption{The tricrittical Casimir force (solid) and the
asymptotic
van der Waals results (dashed) in SES geometry of Fig.\,1
with the same parameters as in Fig.\,2.}
\end{figure}

Eventually, the results for the SES geometry are displayed in Fig.\,3.
For the nonretarded van der Waals force the two asymptotic power laws for
small and large distances (dashed lines)
are taken from Ref.\,\cite{israel}.
The solid line shows the result of Eq.\,(\ref{SES}), again
for $T_c=300\,$K, $\Omega_0=2$, and
$R=1\,\mu$m. 
For $D\gg R$ the curves
approach zero as $\sim D^{-6}$ (van der Waals) and $\sim D^{-2}$ (Casimir),
respectively. 
Again, in both cases the force behaves as $\sim D^{-2}$ for
short distances.
In particular, this means that $f_{\mbox{\tiny SES}}^{tc}$ has
the same $D$-dependence in both limits for $D\ll R$ and $D\gg R$,
with different amplitudes, however, as indicated by the swerve
in Fig.\,3 at the crossover distance $D\simeq R=1\,\mu$m.

\section{Prospects for Experiments}
Probably the best candidates for experiments on the tricritical
Casimir effect are ternary 
or quaternary mixtures of fluids, the latter with the advantage that
they can be ``tuned'' to have a tricritical point at atmospheric pressure \cite{kahlweit}. 
The order parameter $\Phi$ in these systems is Ising-like,
a linear combination of
concentrations of the single components. As said in the Introduction,
the surface will generically favor one of the components
of the fluid, amounting to an explicit symmetry breaking at the surface.
In the scaling limit this leads to the scenario of the extraordinary
transition\cite{binder,bray}. Additionally, systems can be chosen
with their $T_c$ at or around the room temperature. Hence,
as the singular part of the free energy behaves as $\sim T$,
the magnitude of the Casimir force in fluid mixtures should
be larger by about two orders of magnitude than in Helium mixtures
(to be discussed in the next paragraph).

In mixtures of $^3$He and $^4$He the tricritical point
is located at $T_c=867$mK and $x_c=0.675$ mole fraction
of $^3$He.  
It is well known that pure $^4$He 
belongs to the universality class of the ordinary transition,
because the quantum-mechanical wave function that describes
the superfluid state goes to zero at the surface. 
Now, as long as the fraction of $^3$He is
not too large ($x \lesssim  0.54$) also in He mixtures
the scenario of the ordinary transition holds \cite{leibler}.
However, for $x$ beyond a certain value $x_s\simeq 0.54$ 
a superfluid layer (rich in $^4$He) forms already
{\it above} the $\lambda$ line \cite{leibler}, i.e., in the
language of surface critical phenomena a
surface transition \cite{binder} takes place. 
Since the superfluid layer is effectively two-dimensional
and the system possesses a continuous $O(2)$ symmetry, it 
is in a Kosterlitz-Thouless phase \cite{Kothou}. The latter
was also verified by the experiment \cite{McQueeney}.
Note that there is no physical $h_1$ in this system
that couples to the order parameter. 
In the framework of the continuum description the
surface transition is triggered by the negative surface
enhancement \cite{Diehl86}.  

Upon approaching the $\lambda$-line from
above in the concentration range $x_s < x\le x_c$, due to the
growing correlation length 
the superfluid layer becomes thick and a dimensional crossover
\cite{OConnor} from the Kosterlitz-Thouless phase to a three-dimensional 
superfluid phase should take place \cite{Dietrich}.
As a consequence, Helium mixtures 
in the mentioned concentration range also appear to be a candidate
for a system with symmetry-breaking boundary conditions.
However, since the dimensional crossover, in particular the
one between a Kosterlitz-Thouless phase (in $d=2$) and
the ordered phase of the $XY$-model (in $d=3$), is not
fully understood, it is far from clear that the scenario of the
extraordinary transition applies \cite{leibler}. However, concerning the
Casimir force and other critical surface effects there should  
definitely be a difference  
between $x<x_s $ and
$x>x_s$, presumably with a qualitatively new type of Casimir
amplitude for $x>x_s$ that can not be adequately
described in the framework of the known surface universality classes
(ordinary, extraordinary, or special).

An interesting question that could also be addressed in
such experiments is related to critical dynamics. 
The results for the Casimir force presented in the previous sections are
ensemble averages. Due to fluctuations in space and time, 
individual members of an ensemble
deviate from the thermal average, i.e., for instance
in the PP geometry the local pressure will vary in lateral
direction.
Also the van der Waals force, being closely related to the original
electromagnetic Casimir effect, is an average over fluctuations,
this time of quantum-mechanical nature, however.
The typical time scale on which fluctuations of the van der Waals force
take place is given by $D/c$, where $c$ is the speed of light.
On a presumably much larger time scale, behaving as 
$\sim D^{\zeta}$,
the fluctuations of the Casimir force take place, where
$\zeta$ denotes the dynamic
(equilibrium) exponent of model H in the terminology
of Ref.\,\cite{hoha}.
In the PP geometry when $M\gg\xi\gg D$ this time dependence
would be probably absent because the average 
over the lateral directions is taken. In the SPW geometry, however, 
the time dependence of the critical fluctuations taking place on a potentially
macroscopic scale should be observable.

Finally we should like to emphasize the long-range nature of the
Casimir forces specific to symmetry-breaking boundary conditions.
For the critical case it was pointed out by de Gennes \cite{deG80}
and studied in more detail in Refs.\,\cite{B+E95,eisri}
that in the SES geometry for large distances the Casimir free energy
decays as $\sim D^{-2\beta/\nu}$. With $\beta/\nu\simeq 0.52$
for the $d=3$ Ising system, this is already very close to a Coulomb potential.
The tricritical Casimir free energy
falls off {\it exactly} as $\sim D^{-1}$.
As already discussed in Ref.\,\cite{eisri}, this
should have consequences for the thermodynamics of charged
stabilized colloids when the correlation length in the
solvent becomes comparable or larger than the average
distance of colloid particles. In particular we expect
reversible flocculation near a critical or tricritical point.
Flocculation phenomena in fluid mixtures near critical
points have 
already been reported in the literature\cite{Beysens,Kumar}.

\section{Summary and Concluding Remarks}
We studied the Casimir effect at tricritical points
with symmetry-breaking boundary conditions. The case of symmetry-preserving
boundaries was treated earlier in the literature by Krech and Dietrich \cite{Krech91}.
The leading asymptotic behavior of the Casimir force in $d=3$
is determined by the mean-field theory, improved
by renormalization-group considerations. Different from critical systems and
tricritical systems with symmetry-preserving boundaries, where
the Casimir amplitude is {\it universal}, at
a tricritical point with symmetry-breaking boundaries a
{\it nonuniversal} factor occurs. As expressed in (\ref{omega}),
this factor depends on the ratio between a typical macroscopic and 
a microscopic length,
and we estimated it to lie between 2 and 3 for a typical
experimental setting. 

Our calculations were restricted to
bulk tricriticality, but
the results derived
should hold more generally 
when the correlation length $\xi$ is larger or much larger
than $D$. In case that $\xi$ becomes comparable to $D$, the
force should decrease, and in particular for $\xi\ll D$ it should
decay exponentially.
     
Up to date neither the critical nor the tricritical Casimir
effect have been verified in the experiment.
The most promising candidates for experiments on the tricritical
Casimir effect are ternary mixtures and quaternary
mixtures of fluids. Concerning the geometry
we suggested either the parallel-plate geometry, in which
the electromagnetic Casimir effect was verified successfully \cite{lamo},
or the geometry of a sphere near a planar wall, in which the van der
Waals force was measured with an atomic force microscope \cite{Ducker91}.
In particular in fluid mixtures 
the tricritical Casimir effect should dominate
the van der Waals force.

As another candidate for experiments on the tricritical Casimir effect
we discussed Helium mixtures. In this case it turned out
that above a certain concentration
$x_s> 0.54$ and in particular at the tricritical value
$x_c=0.675$ the surface effects are
neither covered by the scenario of the ordinary transition
nor by the one of the extraordinary transition, for
the surface undergoes a transition to
a Kosterlitz-Thouless phase \cite{leibler}.
Hence, it requires further theoretical studies
to obtain the corresponding Casimir amplitudes.

Finally we mention that there are other
obvious directions in which the
theoretical work on the statistical Casimir effect should be extended.
Besides critical and tricritical points, complex fluid mixtures
exhibit a variety of interesting critical phenomena, like 
for example double or quadruple critical points, discussed
in the literature under the heading of {\it reentrant phase
transitions} \cite{Nara}. The parameters in these
systems, the temperature, the pressure, and the various concentrations
can be well controlled, and the experiments reveal an intriguing
spectrum of phenomena as for example the doubling of
critical exponents at a double critical point \cite{Kumar}.
It would be certainly of interest to study
surface critical phenomena and, especially, the Casimir effect also
at these special critical points.\\[4mm]
{\small {\bf Acknowledgements}: We thank H. W. Diehl
for a helpful discussion in the early stage of this work and especially
E. Eisenriegler for the
critical reading of the manuscript, helpful comments, and
hints to the literature.
This work was supported in part by the Deutsche Forschungsgemeinschaft
through Sonderforschungsbereich 237.}


\begin{thebibliography}{99}
%
\bibitem{Casimir48} H. B. G. Casimir, Proc. K. Ned. Akad. Wet. {\bf
51}, 793 (1948); for a recent review on the Casimir effect in QED,
see G. Plunien,
B. M\"uller, and W. Greiner, Phys. Rep. {\bf 134} (1986) 87.
%
\bibitem{lamo} S. Lamoreaux, Phys. Rev. Lett. {\bf 78} (1997) 5.
%
\bibitem{fidege} M. E. Fisher and P.-G. de Gennes, C. R.  Acad. Sci.
Ser. B {\bf 287} (1978) 207.
%
\bibitem{Bloete86} H. W. J. Bl\"ote, J. L. Cardy, and
M. P. Nightingale, Phys. Rev. Lett. {\bf 56} (1986) 742.
%
\bibitem{Affleck} I.\ Affleck, Phys.\ Rev.\ Lett.\ {\bf 56}
(1986) 746.
%
\bibitem{Cardy86} J.\ L.\ Cardy, Nucl. Phys. B {\bf 275} (1986) 200.
%
\bibitem{BuXu91} T. W. Burkhardt and T. Xue, Phys. Rev. Lett. {\bf
66} (1991) 895; Nucl. Phys. {\bf B345} (1991) 653.
%
\bibitem{B+E94} T. W. Burkhardt and E. Eisenriegler, Nucl.\ Phys.\ B
{\bf 424}
[FS] (1994) 487.
%
\bibitem{Symanzik81} K. Symanzik, Nucl. Phys. {\bf B190}, [FS3] (1981) 1.
%
\bibitem{InNi86} M. P. Nightingale and J. O. Indekeu,
Phys. Rev. Lett. {\bf 54} (1985) 1824;
J. O. Indekeu, M. P. Nightingale, and W. V. Wang,
Phys. Rev. B {\bf 34} (1986) 330.
%
\bibitem{Krech91} M. Krech and S. Dietrich, Phys. Rev. Lett. {\bf 66} (1991)
345; {\bf 67} (1991) 1055; Phys. Rev. A {\bf 46} (1992) 1886;
{\bf 46} (1992) 1922.
%
\bibitem{EiSt} E. Eisenriegler and M. Stapper, Phys. Rev. B {\bf 50}
(1994) 10009.
%
\bibitem{B+E95}  T. W. Burkhardt and E. Eisenriegler, Phys. Rev. Lett.
{\bf 74} (1995) 3189.
%
\bibitem{eisri} E. Eisenriegler and U. Ritschel, Phys. Rev. B {\bf 51} (1995) 13717.
%
\bibitem{monte} K. K. Mon, Phys. Rev. Lett. {\bf 54} (1985) 2671;
M. Krech and D. P. Landau, Phys. Rev. E {\bf 53} (1996) 4414.
%
\bibitem{Krech94} M. Krech, {\it The Casimir Effect in Critical
Systems}
(World Scientific, Singapore 1994)
%
\bibitem{binder} K. Binder, 
in {\em Phase Transitions and Critical Phenomena}, Vol.\ 8,
C.\ Domb and J.\ L.\ Lebowitz, eds.\ (London, Academic Press, 1983).
%
\bibitem{bray} A. J. Bray and M. A. Moore, J. Phys. A {\bf 10}
(1977) 1927.
%
\bibitem{lawsa} I. D. Lawrie and S. Sarbach, 
in {\em Phase Transitions and Critical Phenomena}, Vol.\ 9,
C.\ Domb and J.\ L.\ Lebowitz, eds.\ (London, Academic Press, 1984).
%
\bibitem{WegRie} F. J. Wegner and E. K. Riedel, Phys. Rev. Lett,
{\bf 7} (1973) 248.
%
\bibitem{Ducker91} W. A. Ducker, T. J. Senden, and R. M. Pashley,
Nature {\bf 353} (1991) 239.
%
\bibitem{dupla} B. Duplantier, J. Physique (France) {\bf 43} (1982) 991.
%
\bibitem{CardyRev} J. L. Cardy, in {\em Phase Transitions and Critical
Phenomena} Vol. 11,
Eds. C. Domb and J. L. Lebowitz (Academic Press, London, 1987).
%
\bibitem{G+R} S. Gnutzmann and U. Ritschel, Z. Phys. B {\bf 96}
(1995) 391.
%
\bibitem{Diehl86} H. W. Diehl, in {\em Phase Transitions and Critical
Phenomena} Vol\ 10,
Eds. C. Domb and J. L. Lebowitz (Academic Press, London, 1986)
%
\bibitem{Brown80} L. S. Brown, Ann. Phys. (NY) {\bf 126} (1980) 135.
%
\bibitem{ZinnJust} J. Zinn-Justin, {\it Quantum Field Theory and
Critical Phenomena}, (Clarendon Press, Oxford, 1989).
%
\bibitem{eisdi} E. Eisenriegler and H. W. Diehl, Phys. Rev. D {\bf 37}
(1988) 5257.
%
\bibitem{Leach} W. Sarlet and L. J. Bahar, Int. J. Non-Linear Mech.
{\bf 15} (1980) 133; P. G. Leach, J. Math. Phys. {\bf 26} (1985) 2510.
%
\bibitem{Krech97} M. Krech, to be published
%
\bibitem{CardyPaper} J. L. Cardy, Nucl. Phys. B {\bf 240} [FS12] (1984) 514.
%
\bibitem{Gradst} I. S. Gradshteyn and I. M. Ryzhik, {\it Table of
Integrals, Series, and Products}, (Academic Press, New York, 1980).
%
\bibitem{israel} J. N. Israelachvili, {\it Intermolecular and surface forces}, (London, Academic Press, 1991).
%
\bibitem{leibler} S. Leibler and L. Peliti, Phys. Rev. B {\bf 29} (1984)
1253; L. Peliti and S. Leibler, J. Physique Lett. {\bf 45} (1984)
L591.
%
\bibitem{Kothou} J. M. Kosterlitz and D. J. Thouless, J. Phys. C {\bf 6} (1973)
1181.
%
\bibitem{McQueeney} D. McQueeney, G. Agnolet, and J. D. Reppy, Phys. Rev. Lett.
{\bf 52} (1984) 1325.
%
\bibitem{brezin} E. Brezin, J. C. Le Guillou, J. Zinn-Justin, in {\em Phase Transitions and Critical
Phenomena} Vol\ 6,
Eds. C. Domb and J. L. Lebowitz (Academic Press, London, 1984)
%
\bibitem{kahlweit} M. Kahlweit, R. Strey, M. Aratono, M. G. Busse, J. Jen, K. V. Schubert,
J. Chem. Phys. {\bf 95} (1991) 2842.
%
\bibitem{OConnor} D. J. O'Connor and C. R. Stephens, Phys. Rev. Lett. {\bf 72} (1974) 506.
%
\bibitem{Dietrich} S. Dietrich, in {\em Phase Transitions and Critical
Phenomena} Vol\ 12,
Eds. C. Domb and J. L. Lebowitz (Academic Press, London, 1991)
%
\bibitem{hoha} P.\ C.\ Hohenberg and B.\ I.\ Halperin,
Rev.\ Mod.\ Phys.\ {\bf 49} (1977) 435.
%
\bibitem{deG80} P.-G. de Gennes, C. R.  Acad. Sci.
Ser. B {\bf 292} (1981) 701.
%
\bibitem{Beysens} D. Beysens and S. Leibler, J. Physique. Lett.
{\bf 43} (1982) L133;
D. Beysens and D. Est\`eve, Phys. Rev. Lett. {\bf 54} (1985) 2123;
D.\ Beysens, J.-M. Petit, P.\ Narayanan, A.\ Kumar,
and M.\ L.\ Broide, Ber. Bunsen Ges. Phys. Chem. {\bf 98}
(1994) 382.
%
\bibitem{Nara} See P. Narayanan
and A. Kumar, Phys. Rep. {\bf 249} (1994) 135 for an extensive review on this
topic.
%
\bibitem{Kumar} A. Kumar,
Ind. J. Pure Appl. Phys. {\bf 34 }(1996) 742.
%
\end{thebibliography}
\end{document}